\documentclass[12pt]{article}
\usepackage{amsmath}
\usepackage{a4wide,graphicx,psfrag,rotating,feynarts}
\usepackage{epsfig}

\allowdisplaybreaks[1]
\def\s#1{\setbox0=\hbox{$#1$}%
  \rlap{\ifdim\wd0>.7em\kern.22\wd0\else\kern.1\wd0\fi /}#1}

\newcommand{\romd}{{\rm d}}
\newcommand{\hs}{\hat{s}}
\newcommand{\hatt}{\hat{t}}
\newcommand{\ptg}{p^\gamma_{\rm T}}
\newcommand{\ptc}{p^{\gamma,c}_{\rm T}}
\newcommand{\ptcs}{(p^{\gamma,c}_{\rm T})^2}
\newcommand{\beq}{\begin{equation}}
\newcommand{\eeq}{\end{equation}}

\begin{document}

\thispagestyle{empty}
\setcounter{page}{0}
\def\thefootnote{\fnsymbol{footnote}}

{\textwidth 15cm

\begin{flushright}
MPP-2004-25\\
PSI-PR-04-03\\
hep-ph/0402281 \\
\end{flushright}

\vspace{2cm}

\begin{center}

{\Large\sc {\bf Electroweak corrections to 
                \boldmath{$\gamma Z$} production\\[0.3cm]
                at hadron colliders}} 

\vspace{2cm}

{\sc W. Hollik$^a$} and {\sc C. Meier$^{b}$ }

\vspace*{1cm}

     $^a$Max-Planck-Institut f\"ur Physik \\ 
     (Werner-Heisenberg-Institut)\\
     D-80805 M\"unchen, Germany
     
\vspace*{0.4cm}
  
     $^b$Paul-Scherrer Institut \\
     CH-5232 Villigen PSI, Switzerland

\end{center}

\vspace*{2cm}

\begin{abstract}
\noindent
In this paper we present the results from a calculation of the 
full electroweak one-loop corrections 
for  $\gamma Z$ vector-boson pair production
at hadron colliders. 
The cases of proton--antiproton as well as proton--proton
collisions, at the Tevatron and the LHC, respectively,
are considered.
Results are presented for
the distribution of the $\gamma Z$ invariant mass and 
for the transverse momentum of the final-state photon. 
The higher-order electroweak effects are numerically significant,
in particular  for probing possible 
anomalous gauge-boson couplings.
\end{abstract}

}
\def\thefootnote{\arabic{footnote}}
\setcounter{footnote}{0}

\newpage

\section{Introduction}
\label{introduction}
Experiments at  $e^+ e^-$ and hadron colliders 
during the last decade have confirmed the predictions of the
Standard Model to very high accuracy (see \cite{lepewwg} for 
a recent review). In particular,
the interaction of gauge bosons with fermions has been 
tested to a precision at the level of $0.1 \%$.
On the other hand, 
one of the most direct consequences of the Standard Model
as a non-Abelian gauge theory, 
the self couplings of the gauge bosons, 
are known with lower experimental precision.
At the LHC, a substantial improvement in the measurement
of the vector-boson self couplings will become feasible 
via the production of gauge-boson pairs providing the best direct
tests of the non-Abelian gauge symmetry.
Precise predictions from the Standard Model will allow to isolate
possible deviations in the experimental data which may come from
anomalous contributions to the gauge self-interactions
indicating the presence of new physics beyond the Standard Model.
If no deviations are observed, severe bounds on
anomalous vector-boson couplings will be obtained
(see e.g.~\cite{LHC} and references therein).
Production of neutral vector-boson pairs, 
like $\gamma Z$ production, is well suited to search 
for non-zero $ZZ\gamma$ and $Z\gamma\gamma$ couplings, which
vanish in the Standard Model, and could thus provide a clean signal
of new physics.

Previous studies of $\gamma Z$ pair
production in hadronic collisions were first performed at the level of
the Born approximation based on the parton processes 
$q \bar{q} \rightarrow \gamma Z$~\cite{Bij} 
and were later improved
by the NLO QCD corrections~\cite{Ankopp}. The $O(\alpha_s)$
QCD corrections to the hadronic cross section have been shown
to modify the Born result considerably. 
These corrections are of special importance
for high values of the $\gamma Z$ invariant mass
and for high transverse momenta of the photon, i.e.\ in the
regions where also effects from possible
anomalous vector-boson self couplings
are expected to be identified as deviations from the Standard Model
predictions. 
It is therefore important 
to have these predictions under control, which makes the
inclusion of the higher-order contributions a necessity.

Besides the QCD corrections, electroweak higher-order effects
can become significant and have to be taken into account as well.
The one-loop logarithmic electroweak effects in $WZ$ and $W\gamma$
production have been studied in~\cite{accomando}.
Here we consider the hadronic production of  $\gamma Z$ pairs
and present 
a calculation of the full electroweak one-loop corrections
to the parton processes
$q \bar{q} \rightarrow \gamma Z$ and to the hadronic cross
section and distributions. Applying appropriate kinematical cuts, 
numerical results are given for
the invariant-mass distribution of $\gamma Z$ and for the photon  
transverse momentum, 
in $PP$ collisions at the LHC as
well as in $P\bar{P}$ collisions at the Tevatron.

\section{Parton processes}
\label{cs}
\begin{figure}[t]
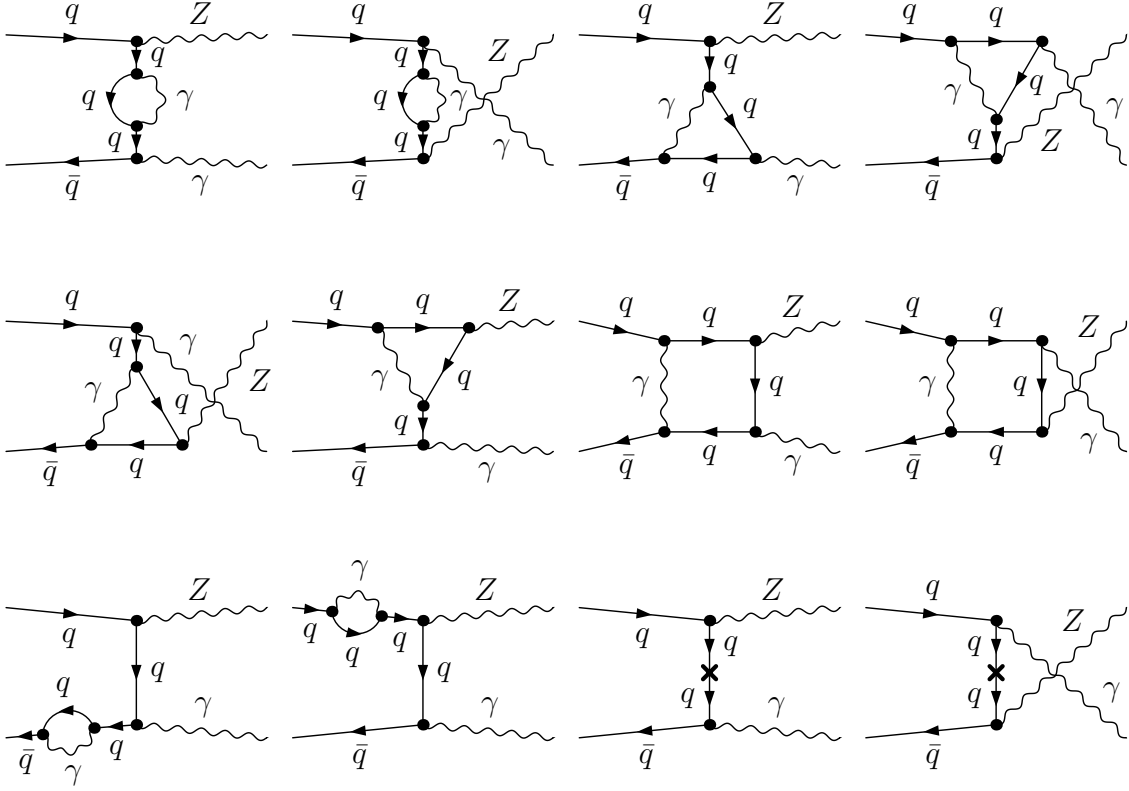

\unitlength=1bp%
\vspace{-3cm}
\begin{feynartspicture}(432,504)(4,3)
\FADiagram{}
\FAProp(0.,15.)(10.,14.5)(0.,){/Straight}{1}
\FALabel(5.0774,15.8181)[b]{$q$}
\FAProp(0.,5.)(10.,5.5)(0.,){/Straight}{-1}
\FALabel(5.0774,4.18193)[t]{$\bar{q}$}
\FAProp(20.,15.)(10.,14.5)(0.,){/Sine}{0}
\FALabel(14.9226,15.8181)[b]{$Z$}
\FAProp(20.,5.)(10.,5.5)(0.,){/Sine}{0}
\FALabel(14.9226,4.18193)[t]{$\gamma$}
\FAProp(10.,14.5)(10.,12.)(0.,){/Straight}{1}
\FALabel(11.07,13.25)[l]{$q$}
\FAProp(10.,5.5)(10.,8.)(0.,){/Straight}{-1}
\FALabel(8.93,6.75)[r]{$q$}
\FAProp(10.,12.)(10.,8.)(1.,){/Straight}{1}
\FALabel(6.93,10.)[r]{$q$}
\FAProp(10.,12.)(10.,8.)(-1.,){/Sine}{0}
\FALabel(13.07,10.)[l]{$\gamma$}
\FAVert(10.,14.5){0}
\FAVert(10.,5.5){0}
\FAVert(10.,12.){0}
\FAVert(10.,8.){0}
\FADiagram{}
\FAProp(0.,15.)(10.,14.5)(0.,){/Straight}{1}
\FALabel(5.0774,15.8181)[b]{$q$}
\FAProp(0.,5.)(10.,5.5)(0.,){/Straight}{-1}
\FALabel(5.0774,4.18193)[t]{$\bar{q}$}
\FAProp(20.,15.)(10.,5.5)(0.,){/Sine}{0}
\FALabel(16.7264,12.9906)[br]{$Z$}
\FAProp(20.,5.)(10.,14.5)(0.,){/Sine}{0}
\FALabel(16.7264,7.00942)[tr]{$\gamma$}
\FAProp(10.,14.5)(10.,12.)(0.,){/Straight}{1}
\FALabel(8.93,13.25)[r]{$q$}
\FAProp(10.,5.5)(10.,8.)(0.,){/Straight}{-1}
\FALabel(8.93,6.75)[r]{$q$}
\FAProp(10.,12.)(10.,8.)(0.8,){/Straight}{1}
\FALabel(7.33,10.)[r]{$q$}
\FAProp(10.,12.)(10.,8.)(-0.75,){/Sine}{0}
\FALabel(12.0716,10.)[l]{$\gamma$}
\FAVert(10.,14.5){0}
\FAVert(10.,5.5){0}
\FAVert(10.,12.){0}
\FAVert(10.,8.){0}
\FADiagram{}
\FAProp(0.,15.)(10.,14.5)(0.,){/Straight}{1}
\FALabel(5.0774,15.8181)[b]{$q$}
\FAProp(0.,5.)(6.5,5.5)(0.,){/Straight}{-1}
\FALabel(3.36888,4.18457)[t]{$\bar{q}$}
\FAProp(20.,15.)(10.,14.5)(0.,){/Sine}{0}
\FALabel(14.9226,15.8181)[b]{$Z$}
\FAProp(20.,5.)(13.5,5.5)(0.,){/Sine}{0}
\FALabel(16.6311,4.18457)[t]{$\gamma$}
\FAProp(10.,14.5)(10.,11.)(0.,){/Straight}{1}
\FALabel(11.07,12.75)[l]{$q$}
\FAProp(6.5,5.5)(13.5,5.5)(0.,){/Straight}{-1}
\FALabel(10.,4.43)[t]{$q$}
\FAProp(6.5,5.5)(10.,11.)(0.,){/Sine}{0}
\FALabel(7.42232,8.60216)[br]{$\gamma$}
\FAProp(13.5,5.5)(10.,11.)(0.,){/Straight}{-1}
\FALabel(12.5777,8.60216)[bl]{$q$}
\FAVert(10.,14.5){0}
\FAVert(6.5,5.5){0}
\FAVert(13.5,5.5){0}
\FAVert(10.,11.){0}
\FADiagram{}
\FAProp(0.,15.)(6.5,14.5)(0.,){/Straight}{1}
\FALabel(3.36888,15.8154)[b]{$q$}
\FAProp(0.,5.)(10.,5.5)(0.,){/Straight}{-1}
\FALabel(5.0774,4.18193)[t]{$\bar{q}$}
\FAProp(20.,15.)(10.,5.5)(0.,){/Sine}{0}
\FALabel(13.3251,7.72981)[tl]{$Z$}
\FAProp(20.,5.)(13.5,14.5)(0.,){/Sine}{0}
\FALabel(18.4482,8.59123)[bl]{$\gamma$}
\FAProp(10.,5.5)(10.,8.5)(0.,){/Straight}{-1}
\FALabel(8.93,7.)[r]{$q$}
\FAProp(6.5,14.5)(13.5,14.5)(0.,){/Straight}{1}
\FALabel(10.,15.57)[b]{$q$}
\FAProp(6.5,14.5)(10.,8.5)(0.,){/Sine}{0}
\FALabel(7.39114,11.199)[tr]{$\gamma$}
\FAProp(13.5,14.5)(10.,8.5)(0.,){/Straight}{1}
\FALabel(12.5341,11.9169)[tl]{$q$}
\FAVert(6.5,14.5){0}
\FAVert(10.,5.5){0}
\FAVert(13.5,14.5){0}
\FAVert(10.,8.5){0}
\FADiagram{}
\FAProp(0.,15.)(10.,14.5)(0.,){/Straight}{1}
\FALabel(5.0774,15.8181)[b]{$q$}
\FAProp(0.,5.)(6.5,5.5)(0.,){/Straight}{-1}
\FALabel(3.36888,4.18457)[t]{$\bar{q}$}
\FAProp(20.,15.)(13.5,5.5)(0.,){/Sine}{0}
\FALabel(18.5763,11.3388)[tl]{$Z$}
\FAProp(20.,5.)(10.,14.5)(0.,){/Sine}{0}
\FALabel(13.2922,12.4512)[bl]{$\gamma$}
\FAProp(10.,14.5)(10.,11.5)(0.,){/Straight}{1}
\FALabel(9.03,13.)[r]{$q$}
\FAProp(6.5,5.5)(13.5,5.5)(0.,){/Straight}{-1}
\FALabel(10.,4.43)[t]{$q$}
\FAProp(6.5,5.5)(10.,11.5)(0.,){/Sine}{0}
\FALabel(7.39114,8.801)[br]{$\gamma$}
\FAProp(13.5,5.5)(10.,11.5)(0.,){/Straight}{-1}
\FALabel(12.9252,7.84012)[bl]{$q$}
\FAVert(10.,14.5){0}
\FAVert(6.5,5.5){0}
\FAVert(13.5,5.5){0}
\FAVert(10.,11.5){0}
\FADiagram{}
\FAProp(0.,15.)(6.5,14.5)(0.,){/Straight}{1}
\FALabel(3.36888,15.8154)[b]{$q$}
\FAProp(0.,5.)(10.,5.5)(0.,){/Straight}{-1}
\FALabel(5.0774,4.18193)[t]{$\bar{q}$}
\FAProp(20.,15.)(13.5,14.5)(0.,){/Sine}{0}
\FALabel(16.6311,15.8154)[b]{$Z$}
\FAProp(20.,5.)(10.,5.5)(0.,){/Sine}{0}
\FALabel(14.9226,4.18193)[t]{$\gamma$}
\FAProp(10.,5.5)(10.,8.5)(0.,){/Straight}{-1}
\FALabel(8.93,7.)[r]{$q$}
\FAProp(6.5,14.5)(13.5,14.5)(0.,){/Straight}{1}
\FALabel(10.,15.57)[b]{$q$}
\FAProp(6.5,14.5)(10.,8.5)(0.,){/Sine}{0}
\FALabel(7.39114,11.199)[tr]{$\gamma$}
\FAProp(13.5,14.5)(10.,8.5)(0.,){/Straight}{1}
\FALabel(12.6089,11.199)[tl]{$q$}
\FAVert(6.5,14.5){0}
\FAVert(10.,5.5){0}
\FAVert(13.5,14.5){0}
\FAVert(10.,8.5){0}
\FADiagram{}
\FAProp(0.,15.)(6.5,13.5)(0.,){/Straight}{1}
\FALabel(3.59853,15.2803)[b]{$q$}
\FAProp(0.,5.)(6.5,6.5)(0.,){/Straight}{-1}
\FALabel(3.59853,4.71969)[t]{$\bar{q}$}
\FAProp(20.,15.)(13.5,13.5)(0.,){/Sine}{0}
\FALabel(16.4015,15.2803)[b]{$Z$}
\FAProp(20.,5.)(13.5,6.5)(0.,){/Sine}{0}
\FALabel(16.4015,4.71969)[t]{$\gamma$}
\FAProp(6.5,13.5)(6.5,6.5)(0.,){/Sine}{0}
\FALabel(5.43,10.)[r]{$\gamma$}
\FAProp(6.5,13.5)(13.5,13.5)(0.,){/Straight}{1}
\FALabel(10.,14.57)[b]{$q$}
\FAProp(6.5,6.5)(13.5,6.5)(0.,){/Straight}{-1}
\FALabel(10.,5.43)[t]{$q$}
\FAProp(13.5,13.5)(13.5,6.5)(0.,){/Straight}{1}
\FALabel(14.57,10.)[l]{$q$}
\FAVert(6.5,13.5){0}
\FAVert(6.5,6.5){0}
\FAVert(13.5,13.5){0}
\FAVert(13.5,6.5){0}
\FADiagram{}
\FAProp(0.,15.)(6.5,13.5)(0.,){/Straight}{1}
\FALabel(3.59853,15.2803)[b]{$q$}
\FAProp(0.,5.)(6.5,6.5)(0.,){/Straight}{-1}
\FALabel(3.59853,4.71969)[t]{$\bar{q}$}
\FAProp(20.,15.)(13.5,6.5)(0.,){/Sine}{0}
\FALabel(17.9814,13.8219)[br]{$Z$}
\FAProp(20.,5.)(13.5,13.5)(0.,){/Sine}{0}
\FALabel(17.8314,6.42814)[tr]{$\gamma$}
\FAProp(6.5,13.5)(6.5,6.5)(0.,){/Sine}{0}
\FALabel(5.43,10.)[r]{$\gamma$}
\FAProp(6.5,13.5)(13.5,13.5)(0.,){/Straight}{1}
\FALabel(10.,14.57)[b]{$q$}
\FAProp(6.5,6.5)(13.5,6.5)(0.,){/Straight}{-1}
\FALabel(10.,5.43)[t]{$q$}
\FAProp(13.5,6.5)(13.5,13.5)(0.,){/Straight}{-1}
\FALabel(12.43,10.)[r]{$q$}
\FAVert(6.5,13.5){0}
\FAVert(6.5,6.5){0}
\FAVert(13.5,6.5){0}
\FAVert(13.5,13.5){0}
\FADiagram{}
\FAProp(0.,15.)(10.,14.)(0.,){/Straight}{1}
\FALabel(4.84577,13.4377)[t]{$q$}
\FAProp(0.,5.)(2.8,5.25)(0.,){/Straight}{-1}
\FALabel(1.53784,4.06114)[t]{$\bar{q}$}
\FAProp(20.,15.)(10.,14.)(0.,){/Sine}{0}
\FALabel(14.8458,15.5623)[b]{$Z$}
\FAProp(20.,5.)(10.,6.)(0.,){/Sine}{0}
\FALabel(15.1542,6.56231)[b]{$\gamma$}
\FAProp(10.,14.)(10.,6.)(0.,){/Straight}{1}
\FALabel(11.07,10.)[l]{$q$}
\FAProp(10.,6.)(6.75,5.75)(0.,){/Straight}{1}
\FALabel(8.49388,4.80957)[t]{$q$}
\FAProp(2.8,5.25)(6.75,5.75)(-0.8,){/Straight}{-1}
\FALabel(4.38035,8.13773)[b]{$q$}
\FAProp(2.8,5.25)(6.75,5.75)(0.8,){/Sine}{0}
\FALabel(5.16965,2.86227)[t]{$\gamma$}
\FAVert(10.,14.){0}
\FAVert(2.8,5.25){0}
\FAVert(10.,6.){0}
\FAVert(6.75,5.75){0}
\FADiagram{}
\FAProp(0.,15.)(3.,14.7)(0.,){/Straight}{1}
\FALabel(1.34577,13.7877)[t]{$q$}
\FAProp(0.,5.)(10.,6.)(0.,){/Straight}{-1}
\FALabel(5.15423,4.43769)[t]{$\bar{q}$}
\FAProp(20.,15.)(10.,14.)(0.,){/Sine}{0}
\FALabel(14.8458,15.5623)[b]{$Z$}
\FAProp(20.,5.)(10.,6.)(0.,){/Sine}{0}
\FALabel(15.1542,6.56231)[b]{$\gamma$}
\FAProp(10.,6.)(10.,14.)(0.,){/Straight}{-1}
\FALabel(11.07,10.)[l]{$q$}
\FAProp(10.,14.)(6.8,14.35)(0.,){/Straight}{-1}
\FALabel(8.23147,13.1142)[t]{$q$}
\FAProp(3.,14.7)(6.8,14.35)(0.8,){/Straight}{1}
\FALabel(4.61784,11.9415)[t]{$q$}
\FAProp(3.,14.7)(6.8,14.35)(-0.8,){/Sine}{0}
\FALabel(5.18216,17.1085)[b]{$\gamma$}
\FAVert(3.,14.7){0}
\FAVert(10.,6.){0}
\FAVert(10.,14.){0}
\FAVert(6.8,14.35){0}
\FADiagram{}
\FAProp(0.,15.)(10.,14.)(0.,){/Straight}{1}
\FALabel(4.84577,13.4377)[t]{$q$}
\FAProp(0.,5.)(10.,6.)(0.,){/Straight}{-1}
\FALabel(5.15423,4.43769)[t]{$\bar{q}$}
\FAProp(20.,15.)(10.,14.)(0.,){/Sine}{0}
\FALabel(14.8458,15.5623)[b]{$Z$}
\FAProp(20.,5.)(10.,6.)(0.,){/Sine}{0}
\FALabel(15.1542,6.56231)[b]{$\gamma$}
\FAProp(10.,10.)(10.,14.)(0.,){/Straight}{-1}
\FALabel(11.07,12.)[l]{$q$}
\FAProp(10.,10.)(10.,6.)(0.,){/Straight}{1}
\FALabel(8.93,8.)[r]{$q$}
\FAVert(10.,14.){0}
\FAVert(10.,6.){0}
\FAVert(10.,10.){1}
\FADiagram{}
\FAProp(0.,15.)(10.,14.)(0.,){/Straight}{1}
\FALabel(5.15423,15.5623)[b]{$q$}
\FAProp(0.,5.)(10.,6.)(0.,){/Straight}{-1}
\FALabel(5.15423,4.43769)[t]{$\bar{q}$}
\FAProp(20.,15.)(10.,6.)(0.,){/Sine}{0}
\FALabel(16.8128,13.2058)[br]{$Z$}
\FAProp(20.,5.)(10.,14.)(0.,){/Sine}{0}
\FALabel(18.1872,7.70582)[bl]{$\gamma$}
\FAProp(10.,10.)(10.,14.)(0.,){/Straight}{-1}
\FALabel(8.93,12.)[r]{$q$}
\FAProp(10.,10.)(10.,6.)(0.,){/Straight}{1}
\FALabel(8.93,8.)[r]{$q$}
\FAVert(10.,14.){0}
\FAVert(10.,6.){0}
\FAVert(10.,10.){1}
\end{feynartspicture}
\vspace{-3cm}
\caption{\label{QEDdiags}One-loop diagrams with virtual photons
           (virtual QED contributions). The cross marks the QED 
           mass-renormalization counter term from the 
           virtual photon contribution to the quark self-energy.}
\end{figure}

At the partonic level, pair-production of neutral vector bosons in lowest
order can only proceed via
quark--antiquark annihilation, $q \bar{q}\rightarrow \gamma Z$
($q=u,d,s,c,b$). 
NLO electroweak corrections consist of the one-loop virtual contributions
and the contributions from real-photon bremsstrahlung .
The subclass of loop diagrams involving
virtual photons gives rise to infrared (IR) divergences; these are cancelled
by including real-photon emission off the quark lines and integrating
over the photon phase space. 
In the NLO-order electroweak contribution to the partonic cross section,
$\romd\sigma^{(1)}_{q \bar{q}}$,
it is convenient to separate the QED corrections from the residual part
of the purely weak corrections, 
\beq
\romd\sigma^{(1)}_{q \bar{q}}=
\romd\sigma^{\rm weak}_{q \bar{q}} +
\romd\sigma^{\rm QED}_{q \bar{q}} \, ,
\label{weakQED}    
\eeq
with the QED part split according to
\beq
\romd\sigma^{\rm QED}_{q \bar{q}} 
= \romd\sigma^{\rm QED}_{q \bar{q},{\rm virt}}
+ \romd\sigma^{\rm QED}_{q \bar{q},{\rm real}} \, .
\label{virtbrems}    
\eeq
Such a separation like in~(\ref{weakQED})
appears natural for neutral-current processes
since the virtual QED corrections form a 
gauge-invariant subset of the full one-loop diagrams; they are
defined as the sum of the diagrams displayed in 
Fig.~\ref{QEDdiags}. Moreover, 
they are UV-finite when quark wave-function renormalization and mass
renormalization, both exclusively from virtual photons, 
are taken into account,
as indicated in the third line of Fig.~\ref{QEDdiags}.

The QED corrections still contain mass-singular
large logarithms of the type $\log\hat{s}/m_q^2$ 
(with the invariant mass squared $\hat{s}$
of the parton process)
arising from real and virtual photons
collinear to a quark or antiquark with mass $m_q$.
These singular terms are universal and can be absorbed by a 
redefinition of the parton distributions functions (PDFs)
of the initial-state quarks.
By this procedure, the mass singularities disappear from the observable
cross section,
and the renormalised PDFs become dependent on the factorization
scale  $\mu_F$, controlled by the Gribov-Lipatov-Altarelli-Parisi (GLAP)
equations~\cite{GLAP}.
Those universal photonic corrections can be taken into account by a 
modifcation of the GLAP equations of QCD introducing an additional
QED-evolution term~\cite{spiesberger}. This leads to small
corrections to the PDFs, at the per-mill level~\cite{LHC}

The consistent treatment of the QED corrections would also require the
inclusion of QED corrections in all the data used for global fitting
of the PDFs. Current determinations of PDFs~\cite{PDF1,PDF2}  
do not include QED corrections, inducing thus an uncertainty  
which, however, should be small compared to the present uncertainties 
on the PDFs.

Absorbing the collinear singularities associated with
initial-state photon radiation into the PDFs introduces 
a QED factorization-scheme dependence.
Our calculation is based on an explicit diagrammatic evaluation,
with the collinear singularities factorized according
to the $\overline{MS}$ scheme~\cite{Baur}, but QED corrections
are not taken into account in the PDF evolution.

For the residual class of 
purely weak corrections, one only encounters loop diagrams 
and counter terms with massive virtual particles. Note
that also in the counter terms for quark mass and wave-function
renormalization all virtual photon contributions have been
separated off as part of the QED corrections.
The counter terms are determined by the electroweak renormalization
scheme, which we choose as the on-shell scheme~\cite{OS1,OS2}.
The partonic cross section derived from the renormalized set of
weak loop diagrams is hence UV-finite 
and does not contain any collinear or infrared divergences.
The practical computation and
evaluation of the complete one-loop terms was done with support of
the packages FeynArts and FormCalc~\cite{FeynArts},
where the version~\cite{OS2} of the on-shell renormalization scheme
is implemented.

\section{Hadronic observables}
\label{integration}

The observable hadronic cross section
for $P+P(\bar{P}) \rightarrow \gamma+Z+ X$,
with a given total hadronic CMS energy $\sqrt{S}$, 
can be written  as a convolution 
of the parton cross sections 
with the corresponding parton luminosities
and summation over the various parton species.
The one-loop order partonic cross section for $q\bar{q}$ annihilation
is obtained from the lowest-order cross section
$\romd \sigma_{q\bar{q}}^{(0)}$
and the NLO contribution~(\ref{weakQED}) after factorization of the
collinear photon contributions in the QED part at the scale $\mu_F$,
\beq  
 \romd \sigma_{q \bar{q}}^{(0+1)}(\hs,\hatt) =
  \romd \sigma_{q\bar{q}}^{(0)}(\hs,\hatt) 
  +  \romd \sigma_{q\bar{q}}^{\rm weak}(\hs,\hatt)
  +  \romd \sigma_{q\bar{q}}^{\rm QED}(\hs,\hatt,\mu_F) \, ,
\label{partonx}
\eeq
with the invariant kinematical variables $\hs$ for the
partonic CMS energy squared and $\hatt$ for the momentum transfer
between $q$ and $\gamma$. 
Integration over $\hatt$ (applying appropriate cuts as described below) 
yields the hadronic cross section in the following way,
\beq
\label{hadronx}
 \sigma^{\gamma Z}_{AB}(S) =
\int_{\tau_0}^{1} \romd \tau \,
\sum_{u\bar{u},d\bar{d},\ldots} 
\frac{ \romd{\cal L}^{A B}_{q \bar q} }{\romd\tau }
\; \sigma_{q\bar{q}}^{(0+1)}(\tau S)
\eeq
with the parton luminosity
\beq
\frac{ \romd{\cal L}^{AB}_{q\bar{q}} }{\romd\tau } =
        \int_{\tau}^{1} \frac{\romd x}{x}  
        \Big[
        q_{A} (x,\mu_F)\, \bar{q}_{B} (\frac{\tau}{x},\mu_F)
        + \bar{q}_{A} (x,\mu_F) \, q_{B} (\frac{\tau}{x},\mu_F)
        \Big]
        \, ,
\eeq
where $q_A(x,\mu_F)$ [$\bar{q}_A(x,\mu_F)$] 
denote the density functions for the quarks [anti-quarks]
in the hadron $A$ carrying a fraction $x$ of the hadron momentum
at the scale $\mu_F$; 
$(A,B)=(P,P)$ for the LHC 
and $(P,\bar{P})$ for the Tevatron.
The lower bound of the $\tau$-integration ($\tau_0$) determines
the minimal invariant mass of the parton system, $\hat s_0 = \tau_0 S$.
In our case, $\tau_0$ depends on the kinematical cuts applied. 
In order to have sufficiently large transverse poton momenta
not too close to the beam axis, 
the following cuts 
\begin{equation} 
 \ptg > \ptc ,
 \quad |y^{\gamma}| < y^{\gamma,c},
\label{cuts}
\end{equation}
are imposed for the photon transverse momentum $\ptg$
and for the pseudo-rapidity $y^\gamma$ 
of the photon, defined by
\begin{equation}
y^{\gamma} = - \log (\tan\frac{\theta}{2})\, ,
\end{equation}
where $\theta$ is the scattering angle in the laboratory frame.
The $\ptg$ cut implies an energy and angular cut in the parton CMS,
\begin{eqnarray}
& &\hs \, > \, \tau_0 S\, =\,  
   \left( \ptc +\sqrt{\ptcs + M_Z^2}\,\; \right)^2 \, , 
                                     \nonumber \\[0.3cm]
& &  |\cos\hat{\theta} | \, <  \, \sqrt{1-
           \frac{4\hs \ptcs}{(\hs-M_Z^2)^2} } \;\;\; .
\end{eqnarray}
In the laboratory frame, $|y^\gamma|$ can still become quite large.
Then, the additional $y^\gamma$ cut restricts photons from coming too 
close to the beam axis in the laboratory frame.
Assignment of specific values for the minimum $\ptg$ and 
maximum $|y^c|$, as used in our numerical analysis,
are contained in Tab.~\ref{cutsvalues}.

Observables of particular interest are the distribution 
for the invariant mass
$M^{\gamma Z}_{\rm inv} = \sqrt{\hs}$
of the $\gamma$--$Z$ final state configuration,
\beq
\frac{\romd \sigma^{\gamma Z}_{AB}}{\romd M^{\gamma Z}_{\rm inv}} \, =\,
\frac{2 M^{\gamma Z}_{\rm inv}}{S} \, \,
\sum_{q,\bar{q}} 
\frac{ \romd{\cal L}^{A B}_{q \bar q} }{\romd\tau }
\; \sigma_{q\bar{q}}^{(0+1)}(\tau S) \, ,
\eeq
and the distribution for the photon transverse momenta,
\beq
\frac{\romd \sigma^{\gamma Z}_{AB}}{\romd \ptg} \, =\,
\int_{\tau_0}^{1} \romd \tau \,
\sum_{q,\bar{q}} 
\frac{ \romd{\cal L}^{A B}_{q \bar q} }{\romd \tau}
\; \frac{\romd \sigma_{q\bar{q}}^{(0+1)}}{\romd \ptg} (\tau S,\ptg) \, .
\eeq

The kinematical constraints arising from 
the cuts divide the phase space into several subvolumes, 
over each of which is integrated separately. 
Integration over phase space was done numerically.
To test the stability of the numerical
integration, the results from an adaptive Gauss algorithm were
checked versus those obtained from the Monte Carlo routine VEGAS7.
As a further check, 
our results for the hadronic cross section in Born
approximation were compared with the results of the 
earlier calculations in~\cite{Bij}.

\section{Numerical evaluation}
\label{results}

\begin{table}[t]

\begin{center}
\begin{tabular}{|c| c| c| c| c| }
\hline
&$\sqrt{S}$ (TeV)&$p_T^{\gamma,c}$ (GeV) &$y^{\gamma,c}$ &$ M_H({\rm GeV})$ \\
\hline
\hline
LHC&$14$&$100$&$2.4$&$115$\\
\hline
Tevatron&$1.8$&$10$&$2.4$&$115$\\
\hline\end{tabular}
\caption{\label{cutsvalues} Input parameters for the
kinematical variables and the Higgs boson mass entering 
the numerical evaluation.} 
\end{center}
\end{table}

The results of the numerical analysis are of lowest order 
with respect to QCD in the parton processes. 
Hence, the numerical values are not yet the
real predictions from the Standard Model; they are given here
to point out and illustrate the effects of the higher-order
electroweak contributions.
The input parameters and the kinematical cuts have been
chosen as listed in Table~\ref{cutsvalues}, together with 
the factorization scale $\mu_F = 2 M_Z$.

QED and weak corrections were calculated separately.
As already mentioned above
for the QED corrections, the collinear singularities were
factorized according to the $\overline{MS}$ scheme, but QED
corrections are not taken into account in the quark distribution
functions used for our analysis~\cite{PDF1}.
The residual non-singular QED corrections are quite small,
at the level of 0.5\%.
There is, however, a left-over QED
factorization-scale dependence, as a consequence of the
missing QED evolution of the PDFs. 
This induces a scale uncertainty
which is of the order of a few per mill as well.
The calculation of the QED contributions thus provides 
more an estimate of the order of magnitude rather than 
precise numbers.  
From a practical point of view, the QED effects for $\gamma Z$
production are negligible since they are small and
covered up by the uncertainties
of the quark distribution functions.

The class of weak corrections induces more significant effects,
in particular in those regions that are of interest for
testing possible anomalous gauge couplings~\cite{LHC}.
Sensitive 
observables for the analysis of anomalous vector-boson couplings are the 
distributions of the hadronic cross sections in terms of the invariant mass 
of $\gamma$ and $Z$ as well as the distribution of the transverse 
momentum of the final state photon.

\begin{center}
\begin{figure}[htbp]

\psfrag{minv}{\begin{tabular}{c}\\ ${\rm M}^{\gamma Z}_{{\rm inv}}$ (GeV)\end{tabular}}
\psfrag{dsigmadminv}{\hspace{-2.9cm}\rotatebox{180}{\begin{tabular}{c}
  {\large $\frac{{\rm d}\sigma^{\gamma
  Z}_{P P}}{{\rm dM}_{\rm inv}^{\gamma Z}}$}(fb/GeV)\\\vspace{0.5cm}\end{tabular}}}
\psfrag{dsigmadminvbar}{\hspace{-2.9cm}\rotatebox{180}{\begin{tabular}{c}
  {\large $\frac{{\rm d}\sigma^{\gamma
  Z}_{P \bar{P}}}{{\rm dM}_{\rm inv}^{\gamma Z}}$}(fb/GeV)\\\vspace{0.5cm}\end{tabular}}}
  \psfrag{0.5}{}
  \psfrag{0.05}{}
  \psfrag{0.005}{}
  \psfrag{1}{\hspace{-1mm}$1$}
  \psfrag{0.1}{\hspace{-0.4cm}$10^{-1}$}
  \psfrag{0.01}{\hspace{-0.25cm}$10^{-2}$}
  \psfrag{0.001}{\hspace{-0.1cm}$10^{-3}$}
  \psfrag{0.0001}{\hspace{0.4cm}$10^{-4}$}
  \psfrag{0}{\hspace{-0.5mm}$0$}
  \psfrag{500}{\hspace{-0.1cm}$500$}
  \psfrag{1000}{\hspace{-0.15cm}$1000$}
  \psfrag{1500}{\hspace{-0.15cm}$1500$}
  \psfrag{2000}{\hspace{-0.15cm}$2000$}
  \psfrag{2500}{\hspace{-0.15cm}$2500$} 
  \psfrag{5}{\hspace{-0.15cm}$5$}
  \psfrag{10}{\hspace{-0.2cm}$10$}
  \psfrag{50}{\hspace{-0.2cm}$50$}
  \psfrag{100}{\hspace{-0.3cm}$100$}  

  \psfrag{S ptc yc}{
    \hspace{-1.75cm}\begin{tabular}{r c l}
   \vspace{0.45cm}&& \\ 
   {\footnotesize $\sqrt{{\rm S}}$}&{\footnotesize $=$}&{\footnotesize$14$ TeV}\\
  {\footnotesize ${\rm p}_{\rm T}^{\gamma,{\rm c}}$}&{\footnotesize $=$}&{\footnotesize $100$
  GeV}\\ 
  {\footnotesize ${\rm y}_{\rm c}$}&{\footnotesize $=$}&{\footnotesize$2.4$}\end{tabular}}

\psfrag{S2 ptc yc}{
    \hspace{-2.0cm}\begin{tabular}{r c l}
   \vspace{0.9cm}&& \\ 
   {\footnotesize $\sqrt{{\rm S}}$}&{\footnotesize $=$}&{\footnotesize$1.8$ TeV}\\
  {\footnotesize ${\rm p}_{\rm T}^{\gamma,{\rm c}}$}&{\footnotesize $=$}&{\footnotesize $10$
  GeV}\\ 
  {\footnotesize ${\rm y}_{\rm c}$}&{\footnotesize
    $=$}&{\footnotesize$2.4$}\end{tabular}}

\hspace{0.5cm}
\epsfig{file=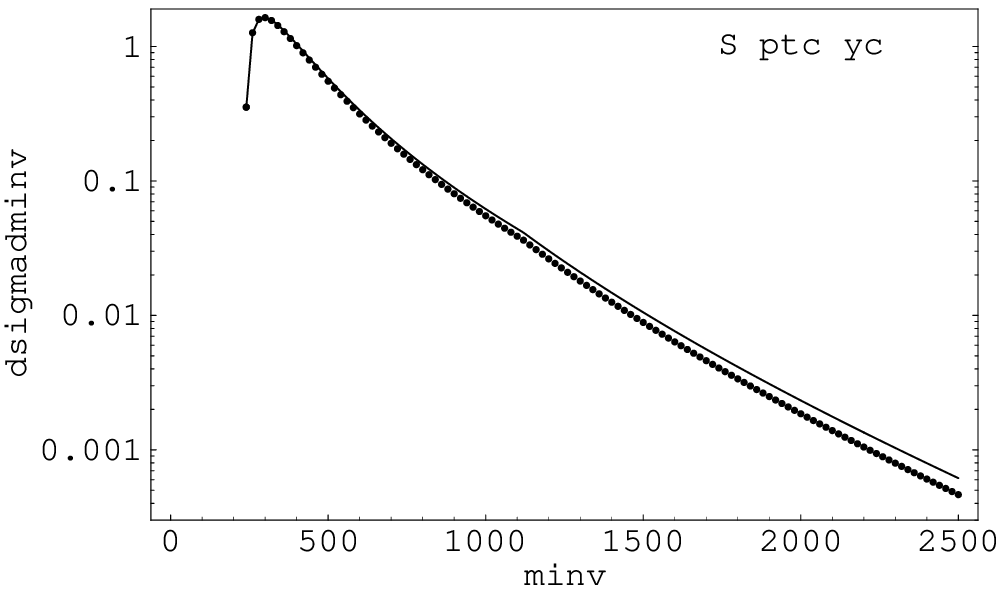,height=6.5cm,width=7cm}
\hspace{1cm}
\psfrag{1000}{\hspace{-0.05cm}$100$}
  \psfrag{2000}{\hspace{-0.05cm}$200$} 
  \psfrag{3000}{\hspace{-0.05cm}$300$}
  \psfrag{4000}{\hspace{-0.05cm}$400$}
  \psfrag{5000}{\hspace{-0.05cm}$500$}
\epsfig{file=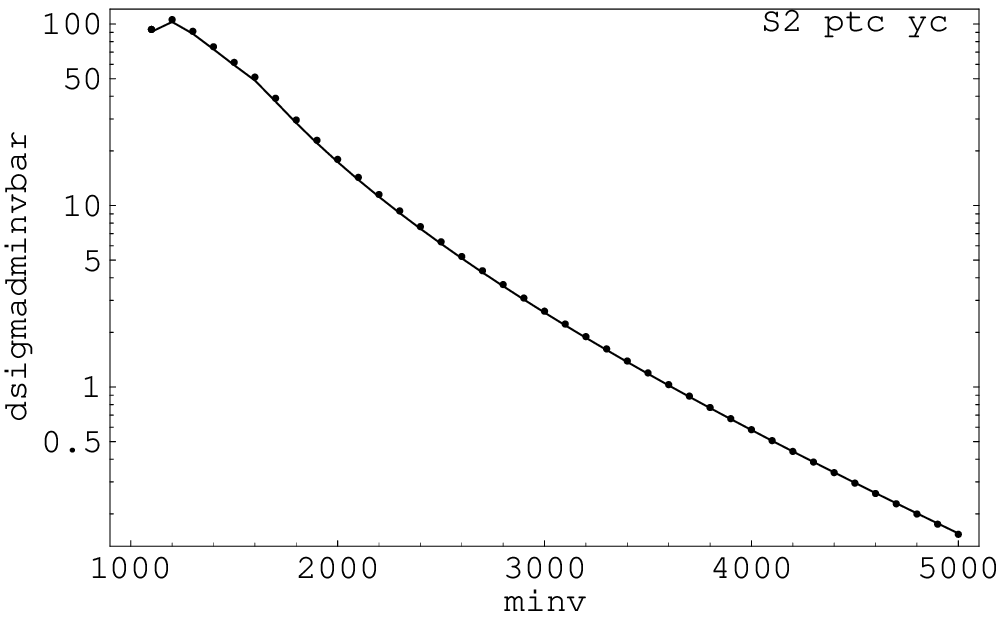,height=6.5cm,width=6.5cm}
\caption{\label{dsdmdiagramm} Distribution of the $\gamma Z$ 
   invariant mass in Born
   approximation (solid line) and including the full 1-loop electroweak
   corrections (dotted line), for the LHC (left case) and for the Tevatron 
  (right case).}
\vspace{-0.5cm}
\end{figure}
\end{center}

\begin{center}
\begin{figure}[htbp]

\psfrag{minv}{\hspace{-0.5cm}\begin{tabular}{c}\\ ${\rm M}^{\gamma
 Z}_{{\rm inv}}$ (GeV)\end{tabular}}
 \psfrag{Deltadsigmadminvweak}{\hspace{-4cm}\rotatebox{180}{\begin{tabular}{c}
  {\large $\Delta \frac{{\rm d}\sigma^{\gamma
  Z,weak}_{P P}}{{\rm dM}_{\rm inv}^{\gamma
 Z}}$}$(\%)$\\\vspace{0.1cm}\end{tabular}}}  

\psfrag{Deltadsigmadminvweakbar}{\hspace{-4cm}\rotatebox{180}{\begin{tabular}{c}
  {\large $\Delta \frac{{\rm d}\sigma^{\gamma
  Z,weak}_{P \bar{P}}}{{\rm dM}_{\rm inv}^{\gamma
 Z}}$}$(\%)$\\\vspace{0.1cm}\end{tabular}}} 

  \psfrag{-0.05}{\hspace{1.5mm}$-5$}
  \psfrag{-0.1}{\hspace{-2mm}$-10$}  
  \psfrag{-0.15}{\hspace{-0.5mm}$-15$}
  \psfrag{-0.2}{\hspace{-2mm}$-20$}
  \psfrag{-0.25}{\hspace{-0.5mm}$-25$}
  
  \psfrag{0.01}{\hspace{3.5mm}$1$}
  \psfrag{0.02}{\hspace{3.5mm}$2$}
  \psfrag{0.03}{\hspace{3.5mm}$3$}
  \psfrag{-0.01}{\hspace{2mm}$-1$}  
  \psfrag{-0.02}{\hspace{1.5mm}$-2$}
  \psfrag{0}{\hspace{-1mm}$0$}
  
  \psfrag{1000}{\hspace{0.15cm}$0$}
  \psfrag{1500}{\hspace{-0.05cm}$500$}
  \psfrag{2000}{\hspace{-0.15cm}$1000$}
  \psfrag{2500}{\hspace{-0.15cm}$1500$} 
  \psfrag{3000}{\hspace{-0.15cm}$2000$}
  \psfrag{3500}{\hspace{-0.15cm}$2500$}
  \psfrag{100}{\hspace{-0.1cm}$100$}
  \psfrag{200}{\hspace{-0.1cm}$200$} 
  \psfrag{300}{\hspace{-0.1cm}$300$}
  \psfrag{400}{\hspace{-0.1cm}$400$}
  \psfrag{500}{\hspace{-0.1cm}$500$}

\psfrag{S ptc yc}{
    \hspace{-1.7cm}\begin{tabular}{r c l}
   \vspace{0.5cm}&& \\ 
   {\footnotesize $\sqrt{{\rm S}}$}&{\footnotesize $=$}&{\footnotesize$14$ TeV}\\
  {\footnotesize ${\rm p}_{\rm T}^{\gamma,{\rm c}}$}&{\footnotesize $=$}&{\footnotesize $100$
  GeV}\\ 
  {\footnotesize ${\rm y}_{\rm c}$}&{\footnotesize $=$}&{\footnotesize$2.4$}\end{tabular}}
  
\psfrag{S2 ptc yc}{
    \hspace{-1.6cm}\begin{tabular}{r c l}
   \vspace{0.35cm}&& \\ 
   {\footnotesize $\sqrt{{\rm S}}$}&{\footnotesize $=$}&{\footnotesize$1.8$ TeV}\\
  {\footnotesize ${\rm p}_{\rm T}^{\gamma,{\rm c}}$}&{\footnotesize $=$}&{\footnotesize $10$
  GeV}\\ 
  {\footnotesize ${\rm y}_{\rm c}$}&{\footnotesize
    $=$}&{\footnotesize$2.4$}\end{tabular}}
\hspace{0.5cm}
\epsfig{file=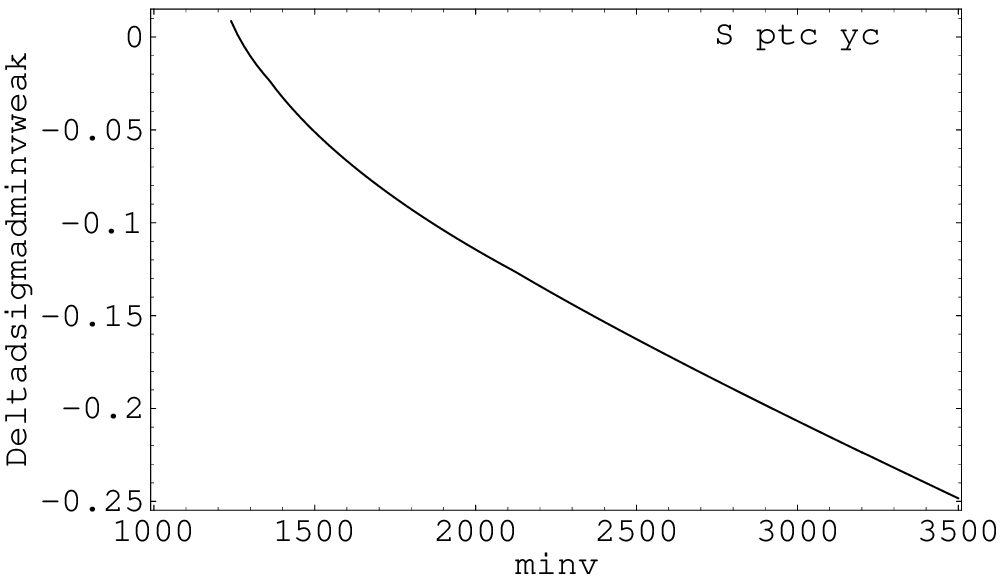,height=6.5cm,width=7cm}
\hspace{1cm}
\epsfig{file=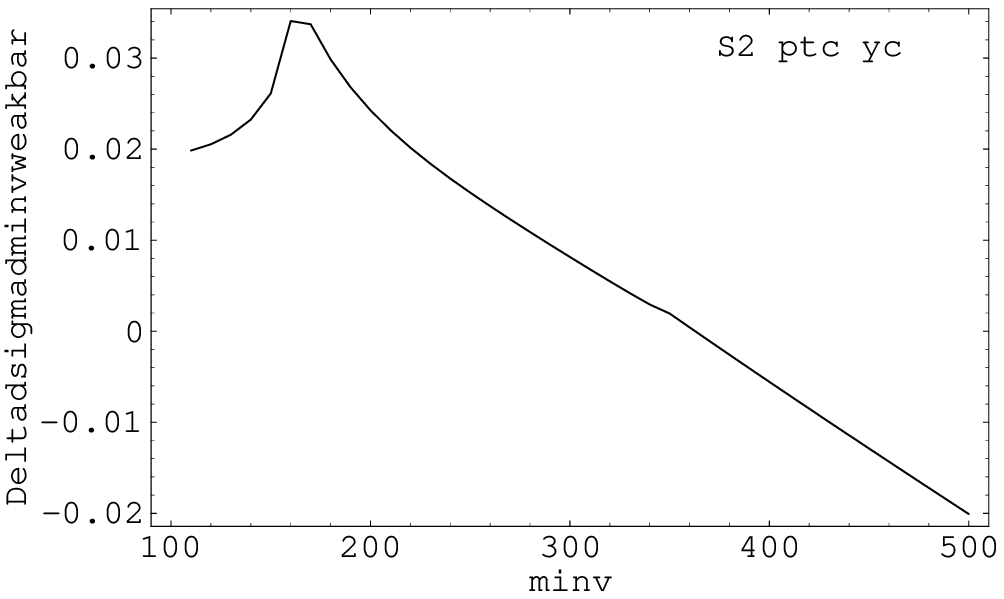,height=6.5cm,width=7cm}
\caption{\label{dsdmdiagrammweak} Weak corrections to the
   distribution of the $\gamma Z$ invariant mass relative to the
   Born result, for the LHC (left case) and for the
   Tevatron (right case).}
\end{figure}
\end{center}

To illustrate the electroweak higher-order effects,
we show in Fig.~\ref{dsdmdiagramm} 
the distribution of the $\gamma Z$ invariant mass in lowest-order 
approximation in comparison to the corresponding result including the 
full 1-loop electroweak corrections, both for the
LHC and for the Tevatron. 
To be more quantitative, 
Fig.~\ref{dsdmdiagrammweak} diplays 
the purely weak corrections to the $M^{\gamma Z}_{\rm inv}$ distribution,
relative to the Born result. 
In the LHC case, 
the weak corrections reach a considerable size around 
$20\%$ for high invariant masses, 
whereas for the Tevatron they are in the lower percentage region.

In a similar way, 
in Figures \ref{dsdptdiagramm} and
\ref{dsdptdiagrammweak}, the corresponding 
results are shown for the $p_{\rm T}$ distribution of the
final-state photon. 
Again, the weak contributions 
are sizeable in case of the LHC, at the level of $30 \%$ and more
for high $\ptg$. 
But also for the Tevatron case, they can exceed the level of $10 \%$.
The origin of the large effects at high invariant masses or
high $p_{\rm T}$, respectively, are large logarithms,
e.g.\ of the type $\alpha \log^2(\hs/M_Z^2)$, in the one-loop 
contributions to the parton processes.

Finally,
the electroweak loop terms depend also on the mass of the Higgs boson,
which was kept at $M_H = 115$ GeV for the numerical results
displayed in the figures. 
The dependence on $M_H$ is not very strong,
yielding a variation of a few per cent (relative) 
of the electroweak corrections setting $M_H = 1$ TeV.

The electroweak corrections are negative whenever they are sizeable,
opposite to the positive NLO QCD corrections 
\cite{Ankopp}, and thus partially compensate the NLO
QCD contributions. \\[0.3cm]

To summarize,
we have presented the results of a full electroweak one-loop
calculation for the production of
$\gamma Z$ pairs at hadron colliders, evaluated 
for both the Tevatron and the LHC. 
Like for other neutral-current processes, the electroweak
$O(\alpha)$ corrections naturally decompose into QED and
weak contributions, which are separately gauge invariant
and UV finite. The QED corrections, arising  
from virtual and real photon emission, contain
collinear singularities which are factorized and absorbed
in the PDFs. The remaining QED corrections are small, below
the per-cent level, and insignificant in view of the uncertainties
of the quark distribution functions.
The  weak corrections, however, are sizeable for large
parton energies or transverse photon momenta, respectively,
especially for the LHC, and have to be taken into account 
together with the NLO QCD corrections for testing the
gauge structure of the Standard Model and probing the
existence of anomalous gauge couplings.

\begin{center}
\begin{figure}[htbp]
\psfrag{pt}{\hspace{-0.5cm}\begin{tabular}{c}\\ ${\rm p}_{\rm T}^{\gamma}$ (GeV)\end{tabular}}
\psfrag{dsigmadpt}{\hspace{-3cm}\rotatebox{180}{\begin{tabular}{c}
  {\large $\frac{{\rm d}\sigma^{\gamma
  Z}_{P P}}{{\rm dp}_{\rm T}^{\gamma
 }}$}(fb/GeV)\\\vspace{0.1cm}\end{tabular}}}
\psfrag{dsigmadptbar}{\hspace{-3cm}\rotatebox{180}{\begin{tabular}{c}
  {\large $\frac{{\rm d}\sigma^{\gamma
  Z}_{P \bar{P}}}{{\rm dp}_{\rm T}^{\gamma
 }}$}(fb/GeV)\\\vspace{0.1cm}\end{tabular}}}   
  \psfrag{0.001}{\hspace{-1mm}$10^{-3}$}
  \psfrag{0.01}{\hspace{-2.5mm}$10^{-2}$}  
  \psfrag{0.1}{\hspace{-4mm}$10^{-1}$}
  \psfrag{1}{\hspace{-0.5mm}$1$}
  \psfrag{10}{\hspace{-1.5mm}$10$}
  \psfrag{100}{\hspace{-2mm}$100$}
  \psfrag{1000}{\hspace{-0.15cm}$1000$}
  \psfrag{0}{\hspace{-0.5mm}$0$}
  \psfrag{200}{\hspace{-0.1cm}$200$}
  \psfrag{300}{\hspace{-0.1cm}$300$}
  \psfrag{400}{\hspace{-0.1cm}$400$}
  \psfrag{600}{\hspace{-0.1cm}$600$}
  \psfrag{600}{\hspace{-0.1cm}$600$}
  \psfrag{800}{\hspace{-0.1cm}$800$}  
  \psfrag{1200}{\hspace{-0.15cm}$1200$}
  \psfrag{10000}{\hspace{0.5mm}$100$}
  \psfrag{20000}{\hspace{0.5mm}$200$}
  \psfrag{30000}{\hspace{0.5mm}$300$}
  \psfrag{40000}{\hspace{0.5mm}$400$}
\psfrag{S ptc yc}{
    \hspace{-2cm}\begin{tabular}{r c l}
   \vspace{-2.2cm}&& \\ 
   {\footnotesize $\sqrt{{\rm S}}$}&{\footnotesize $=$}&{\footnotesize$14$ TeV}\\
  {\footnotesize ${\rm p}_{\rm T}^{\gamma,{\rm c}}$}&{\footnotesize $=$}&{\footnotesize $100$
  GeV}\\ 
  {\footnotesize ${\rm y}_{\rm c}$}&{\footnotesize $=$}&{\footnotesize$2.4$}\end{tabular}}
\psfrag{S2 ptc yc}{
    \hspace{-2cm}\begin{tabular}{r c l}
   \vspace{-1.0cm}&& \\ 
   {\footnotesize $\sqrt{{\rm S}}$}&{\footnotesize $=$}&{\footnotesize$1.8$ TeV}\\
  {\footnotesize ${\rm p}_{\rm T}^{\gamma,{\rm c}}$}&{\footnotesize $=$}&{\footnotesize $10$
  GeV}\\ 
  {\footnotesize ${\rm y}_{\rm c}$}&{\footnotesize
    $=$}&{\footnotesize$2.4$}\end{tabular}}
\hspace{0.5cm}
\epsfig{file=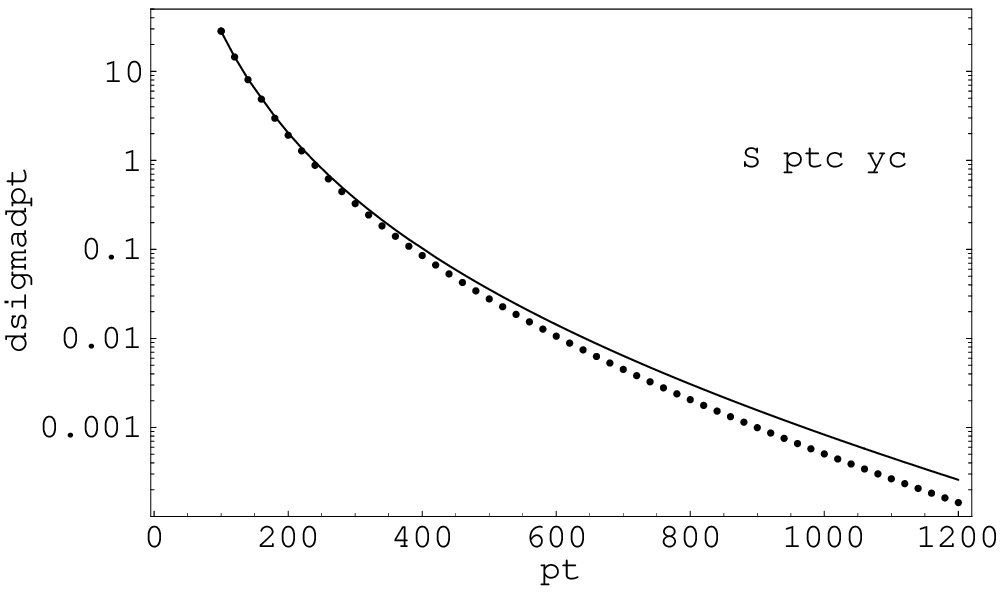,height=6.5cm,width=7cm}
\psfrag{1000}{\hspace{-3mm}$1000$}
\hspace{1cm}
\epsfig{file=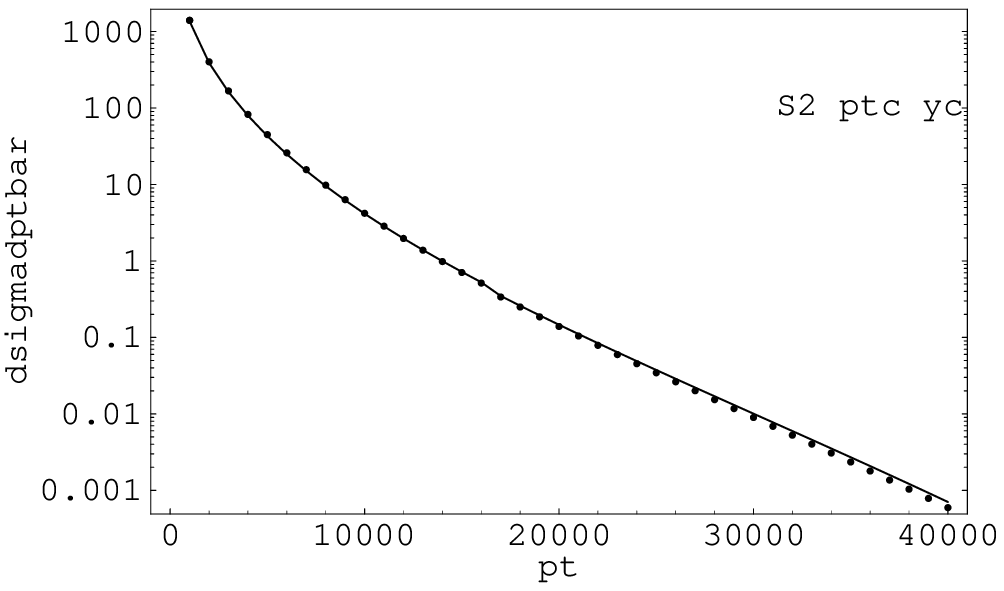,height=6.5cm,width=7cm}
\caption{\label{dsdptdiagramm} Distribution of the transverse momentum in Born
   approximation (solid line) and including the full 1-loop electroweak
   corrections (dotted line) for the LHC (left case) and for the
   Tevatron (right case).}
\end{figure}

\begin{figure}[htbp]
\psfrag{pt}{\hspace{-0.5cm}\begin{tabular}{c}\\ 
 ${\rm p}^{\gamma}_{{\rm T}}$ (GeV)\end{tabular}}
 
\psfrag{Deltadsigmadptweak}{\hspace{-3.5cm}\rotatebox{180}{\begin{tabular}{c}
  {\large $\Delta \frac{{\rm d}\sigma^{\gamma
  ,weak}_{P P}}{{\rm p}_{\rm T}^{\gamma
 Z}}$}$(\%)$\\\vspace{0.3cm}\end{tabular}}}  
 
\psfrag{Deltadsigmadptweakbar}{\hspace{-3.5cm}\rotatebox{180}{\begin{tabular}{c}
  {\large $\Delta \frac{{\rm d}\sigma^{\gamma
  ,weak}_{P \bar{P}}}{{\rm p}_{\rm T}^{\gamma
 Z}}$}$(\%)$\\\vspace{0.3cm}\end{tabular}}}  
  \psfrag{-0.05}{\hspace{1mm}$-5$}
  \psfrag{-0.1}{\hspace{-2.5mm}$-10$}  
  \psfrag{-0.15}{\hspace{-1mm}$-15$}
  \psfrag{-0.2}{\hspace{-3mm}$-20$}
  \psfrag{-0.25}{\hspace{-1mm}$-25$}
  \psfrag{-0.3}{\hspace{-3mm}$-30$}
  \psfrag{-0.4}{\hspace{-3mm}$-40$}
  \psfrag{0}{\hspace{-1.5mm}$0$}

  \psfrag{10000}{\hspace{2mm}$0$}
  \psfrag{10100}{$100$}
  \psfrag{10200}{$200$}
  \psfrag{10300}{$300$}
  \psfrag{10400}{$400$}
  \psfrag{10600}{$600$}
  \psfrag{10800}{$800$}  
  \psfrag{11000}{\hspace{-0.05cm}$1000$}
  \psfrag{11200}{\hspace{-0.05cm}$1200$}
\psfrag{S ptc yc}{
    \hspace{-1.85cm}\begin{tabular}{r c l}
   \vspace{0.85cm}&& \\ 
   {\footnotesize $\sqrt{{\rm S}}$}&{\footnotesize $=$}&{\footnotesize$14$ TeV}\\
  {\footnotesize ${\rm p}_{\rm T}^{\gamma,{\rm c}}$}&{\footnotesize $=$}&{\footnotesize $100$
  GeV}\\ 
  {\footnotesize ${\rm y}_{\rm c}$}&{\footnotesize $=$}&{\footnotesize$2.4$}\end{tabular}}
  
\psfrag{S2 ptc yc}{
    \hspace{-1.2cm}\begin{tabular}{r c l}
   \vspace{0.75cm}&& \\ 
   {\footnotesize $\sqrt{{\rm S}}$}&{\footnotesize $=$}&{\footnotesize$1.8$ TeV}\\
  {\footnotesize ${\rm p}_{\rm T}^{\gamma,{\rm c}}$}&{\footnotesize $=$}&{\footnotesize $10$
  GeV}\\ 
  {\footnotesize ${\rm y}_{\rm c}$}&{\footnotesize
    $=$}&{\footnotesize$2.4$}\end{tabular}}
\vspace{-1cm}
\hspace{0.5cm}
\epsfig{file=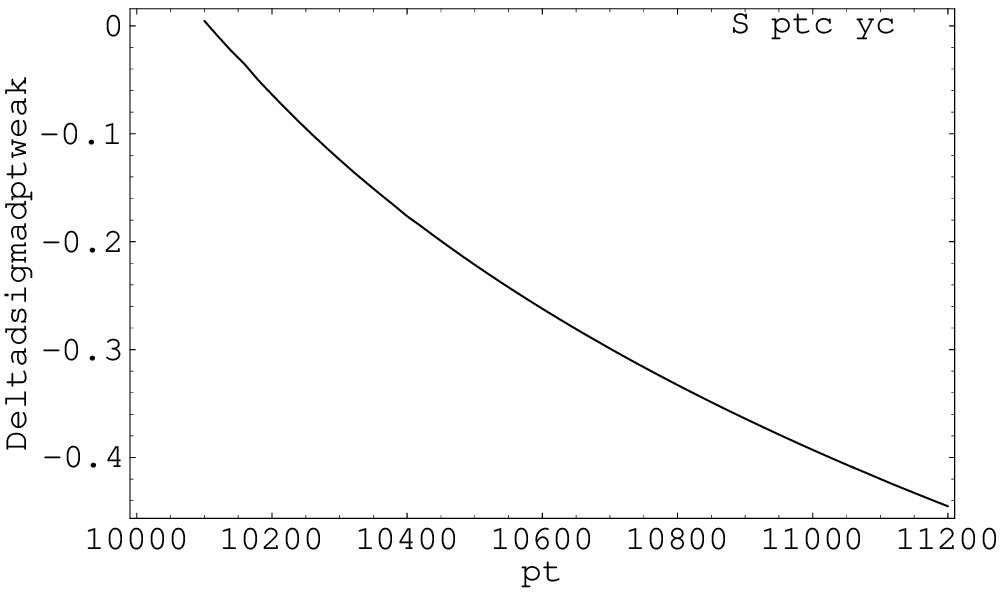,height=6.5cm,width=7cm}
\hspace{1cm}
\epsfig{file=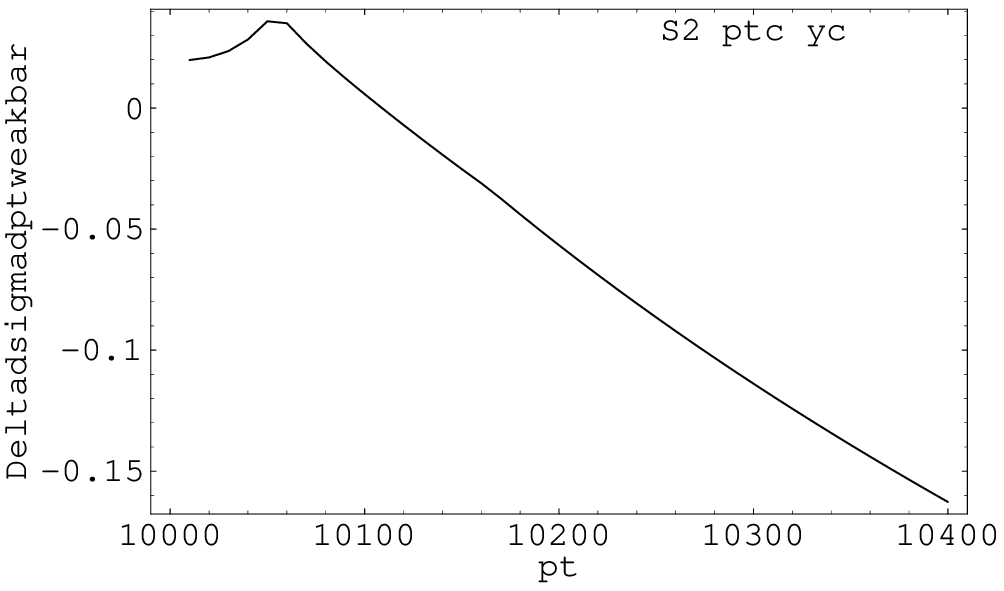,height=6.5cm,width=7cm}
\caption{\label{dsdptdiagrammweak} Weak corrections to the
   distribution of the transverse momentum relative to the
   Born result, for the LHC (left diagram) and for the
   Tevatron (right diagram).}
\end{figure}

\end{center}


\section*{Acknowledgement}
W.H. thanks the Kavli Institut for Theoretical Physics,
where this work was completed.
This research was supported in part by the European Community's
Human Potential Programme under contract HPRN-CT-2000-00149
``Physics at Colliders'' and by the National Science Foundation
under Grant No.\ PHY99-07949.

\pagebreak

\end{document}